\documentclass[twocolumn]{autart}    

\usepackage{graphicx}          

\usepackage[utf8]{inputenc}
\usepackage{psfrag}
\usepackage{amsmath}
\usepackage{amssymb}
\usepackage{bm}
\usepackage{xcolor}	
\usepackage{algorithm}
\usepackage{algpseudocode}

 \usepackage{natbib}

\newcommand{\sP}{\mathbb{P}}
\newcommand{\sN}{\mathcal{N}}
\newcommand{\sV}{\mathcal{V}}
\newcommand{\sE}{\mathcal{E}}
\newcommand{\sF}{\mathcal{F}}

\newcommand{\R}{\mathbb{R}}
\newcommand{\E}{\mathbb{E}}
\newcommand{\Z}{\mathbb{Z}}

\newcommand{\abs}[1]{\left\vert #1 \right\vert}

\newtheorem{theorem}{Theorem}
\newtheorem{lemma}{Lemma}
\newtheorem{proposition}{Proposition}
\newtheorem{definition}{Definition}
\newtheorem{remark}{Remark}

\begin{document}

\begin{frontmatter}

\title{Resilient consensus for multi-agent systems subject to differential privacy requirements\thanksref{footnoteinfo}} 
\thanks[footnoteinfo]{
This work was conducted while the authors were with IBM Research Ireland. Corresponding author: Giovanni Russo.}
%
\author[Napoli]{Davide Fiore}\ead{dvd.fiore@gmail.com},             
\author[Dublin]{Giovanni Russo}\ead{giovanni.russo1@ucd.ie}  
%
%
\address[Napoli]{Department of Electrical Engineering and Information Technology, University of Naples Federico II, Via Claudio 21, 80125 Naples, Italy}  
\address[Dublin]{School of Electrical and Electronic Engineering, University College Dublin, Belfield, Dublin 4, Ireland}


\begin{abstract}                          
We consider multi-agent systems interacting over directed network topologies where a subset of agents is adversary/faulty  and where the non-faulty agents have the goal of reaching consensus, while fulfilling a differential privacy requirement on their initial conditions. To address this problem, we develop an update law for the non-faulty agents. Specifically, we propose a modification of the so-called Mean-Subsequence-Reduced (MSR) algorithm, the Differentially Private MSR (DP-MSR) algorithm, and characterize three important properties of the algorithm: correctness, accuracy and differential privacy. We show that if the network topology is $(2f +1)$-robust, then the algorithm allows the non-faulty agents to reach consensus despite the presence of up to $f$ faulty agents and we characterize the accuracy of the algorithm. Furthermore, we also show in two important cases that our distributed algorithm can be tuned to guarantees differential privacy of the initial conditions and the differential privacy requirement is related to the maximum network degree. The results are illustrated via simulations.
\end{abstract}

\end{frontmatter}
\section{Introduction}
The introduction of low-cost, high performance and connected devices lead, over the last few years, to a paradigm shift on how engineering systems are designed \citep{di2016convergence,albert2000error,fiore2017exploiting}. The philosophy design for those systems is indeed transitioning from a centralized to a distributed paradigm and, in this context, two fundamental challenges are: (i) the fact that agents/nodes might be subject of failures or attacks; and (ii) privacy preservation of individual information. Power grids \citep{Dor_Bul_14} and smart transportation systems \citep{7932942} are examples of systems where control and coordination strategies cannot neglect these fundamental requirements of privacy preservation and fault tolerance. Motivated by this, we present an algorithm that: (i) allows a multi-agent system to reach consensus despite a subset of agents being faulty; and (ii) guarantees differential privacy of the initial conditions of the non-faulty agents. We now briefly review some related works on Differential Privacy and Resilient Consensus.

\noindent {\bf Differential Privacy.} The notion of differential privacy was originally introduced in \citep{dwork2006our,dwork2006calibrating} in relation to the design of privacy preserving mechanisms of databases. From the intuitive viewpoint, a database is differentially private if the removal or addition of a single database entry does not substantially affect the outcome of any analysis on the database data. This privacy definition is also particularly well-suited for networked multi-agent systems, where communication channels might not be fully secure and where agents might not trust each other. In the context of multi-agent systems, protecting initial information of each agent often implies protecting the current state of the network. This is the case, for example, of platoon systems, see e.g. \citep{7937859}, where each vehicle in the platoon wishes to keep private its initial position.  Based on such observations, \citep{huang2012differentially} considered the problem of designing a distributed consensus algorithm for multi-agent systems (evolving over an undirected graph topology) guaranteeing differential privacy of the initial conditions of the agents. In particular, the paper shows that if each agent adds some decaying Laplacian noise to its averaging dynamics, then mean square convergence to a random consensus value, whose expected value depends on the  undirected network topology, can be proved and differential privacy of the initial conditions can be guaranteed. More recently, \citep{nozari2017differentially} proposed an algorithm which guarantees both almost sure convergence to an unbiased estimate of the average of agents' initial conditions and differential privacy of their values, allowing individual agents, interacting over an undirected graph, to independently choose their level of privacy. The concept of differential privacy has been also successfully used in distributed estimation \citep{le2014differentially} and optimization \citep{7039718,Hua_Mit_15,7577745}, while other notions of privacy, see \citep{mo2017privacy} and references therein for further details, have been also investigated for consensus.

\noindent {\bf Resilient Consensus.} Resilient consensus has received much research attention and we refer the reader to \citep{tseng2016note} and \citep{leblanc2013resilient} for a detailed literature review. The original formulation of the problem can be traced back to  \citep{lamport1982byzantine,dolev1982byzantine}. 
In such works, the authors introduced the \emph{Approximate Byzantine Consensus problem} where the agents have to agree on some legit value despite the fact that at most $f$ agents may be adversary (also termed faulty or Byzantine). The networks considered in the paper were complete and, for such networks, the \emph{Mean-Subsequence-Reduced (MSR)} algorithm was proposed, see also \citep{kieckhafer1994reaching}. Tight conditions for approximate consensus using the MSR algorithm in directed networks with $f$ Byzantine agents were derived in \citep{vaidya2012iterative}, while, in \citep{leblanc2013resilient,zhang2012robustness} a modification of the MSR algorithm has been proposed, the Weighted-MSR.  The conditions derived therein were stated in terms of the so-called $r$-robustness property that characterizes the local redundancy of information flow \citep{zhang2015notion}.  Recent works \citep{leblanc2011consensus,leblanc2012low} proposed a continuous-time variation of the MSR algorithm, while in \citep{su2016reaching} the authors allowed agents to take as input not only the information from their neighbors but also the information provided by the agents that are up to $\ell$ \emph{hops} away. In \citep{8100900,dibaji2017resilient}, a modification of the MSR algorithm has been proposed to solve the resilient consensus problem for asynchronous multi-agent systems with quantized communication and delays. The works \citep{sundaram2011distributed,pasqualetti2012consensus} proposed a different approach to detect and identify malicious agents, which is based on observability and reconstruction of the initial conditions. Other complementary approaches to the problem include adding trusted agents \citep{abbas2017improving} and designing self-triggered control strategies \citep{7855691}.
\subsection{Contributions of the paper}
 In this paper we combine differential privacy and resilient consensus in a multi-agent system with directed communication topology. For such systems, we introduce a distributed algorithm guaranteeing that consensus can be attained by honest agents in presence of a certain number of malicious agents along with differential privacy of their initial states. We characterize algorithm correctness and accuracy. Moreover, in two special cases, we show how differential privacy of the initial conditions of the non-faulty agents can be guaranteed. The Differentially Private MSR (DP-MSR) algorithm presented here is the main contribution of this paper and, to the best of our knowledge, this is the first algorithm that simultaneously handles differential privacy and fault tolerance requirements over directed graphs. Indeed: 
 \begin{description}
 \item[(i)] in the literature on differentially private consensus \citep{huang2012differentially,nozari2017differentially,mo2017privacy}, it is assumed that the graph topology is undirected and that there are no malicious agents;
\item[(ii)] vice versa, the literature on resilient consensus (see the related work section) considers agents evolving over directed graphs but, in such works, no differentially private mechanisms are considered.
\end{description}
\section{Mathematical Preliminaries}
Throughout the paper, we denote by $\R$, $\R_{>0}$, $\Z_{\ge 0}$ the set of real numbers, positive real numbers and non-negative integers, respectively. Let $A \in \R^{n\times n}$ ($x\in\R^n$); we denote by $\lVert A\rVert$ ($\lVert x\rVert$) the Euclidean matrix (vector) norm and by $\lVert A\rVert_F=[ \mathrm{tr}(A^T A) ]^{1/2}$ the Frobenius norm ($\mathrm{tr}(\cdot)$ denotes the trace of a square matrix).
Let $\mathcal{A}$ and $\mathcal{B}$ be two sets, then $\abs{\mathcal{A}}$ is the cardinality of $\mathcal{A}$ and we denote by $\mathcal{A}\cup \mathcal{B}$, $\mathcal{A}\cap\mathcal{B}$ and $\mathcal{A}\setminus\mathcal{B}$ the union, intersection and difference of the sets. Let $(\mathbb{R}^n)^\mathbb{N}$ denote the space of vector-valued sequences in $\mathbb{R}^n$. Then, for $\{x(k)\}_{k=0}^\infty\in(\mathbb{R}^n)^\mathbb{N}$, we set $\mathbf{x} := \{x(k)\}_{k=0}^\infty$ and $\mathbf{x}_k := \{x(h)\}_{h=0}^k$. For any topological space, say $X$, we denote by $\mathcal{B}(X)$ the set of Borel subsets of $X$. We denote by $\overline{\mathrm{co}}\{\mathcal{A}\}$ the closed convex hull of the set $\mathcal{A}$. Finally, we denote by $\mathbf{1}$ the column vector of adequate dimension with all entries equal to $1$, and by $I_n$ the identity matrix of size $n$.
\subsection{Probability theory}
Consider a probability space $(\Omega,\Sigma,\sP)$, where $\Omega$ is the sample space, $\Sigma$ is the collection of all the events, and $\sP$ is the probability measure, respectively. 
A random variable is a measurable function $X: \omega\in\Omega\rightarrow \R$, and we denote its expected value by $\E [X]$. Chebyshev's inequality states that, for any $N\in\mathbb{R}_{> 0}$ and any random variable $X$ with finite expected value $\mu$ and finite nonzero variance $\sigma^2$, $\sP\{\abs{X-\mu}\geq N\sigma \}\leq \frac{1}{N^2}$.  We denote by $X\sim \mathrm{Lap}(b)$ a zero mean random variable with Laplace distribution and variance $2b^2$, and with probability density function $\mathcal{L}(x; b)=\frac{1}{2b}e^{-\frac{\abs{x}}{b}}$ for any $x\in\mathbb{R}$.  Finally, consider the random $n$-dimensional vector $X=[X_1,\dots, X_n]^T$. The expected value of $X$ is a vector whose elements are the expected value of elements of $X$, that is $\E[X]=\left[ \E[X_1], \dots , \E[X_n]\right]^T$, and the covariance matrix of $X$ is $\mathrm{Cov}(X)=\E [ (X-\E[X])(X-\E[X])^T ]$. We denote with $\mathrm{var}(X)$ the trace of the covariance matrix. That is, $\mathrm{var}(X)=\mathrm{tr}\left( \mathrm{Cov}(X) \right)=\E[X^TX]-\E[X]^T \E[X]$.
\subsection{Directed graphs and $r$-robustness}
We consider agents interacting over a network topology modeled as a directed graph, or \emph{digraph}. We denote a digraph by $\mathcal{G}(\mathcal{V},\mathcal{E})$, where $\mathcal{V}$ is the \emph{node set}, and $\mathcal{E}\subseteq\mathcal{V}\times \mathcal{V}$ is the \emph{directed edge set}. We denote the number of nodes in the graph by $n$ and we consider graphs for which $n \ge 2$. A directed edge is an ordered pair of distinct vertices and we denote by $(j,i)\in\mathcal{E}$ an edge from node $j\in\mathcal{V}$ to node $i\in\mathcal{V}$. In the context of this paper, the directed edge $(j,i)$ models the information flow from agent $j$ to agent $i$. The set of \emph{in-neighbors} of node $i$ is denoted by $\mathcal{N}_i :=\{j\in\mathcal{V}:(j,i)\in\mathcal{E} \}$, and the \emph{in-degree} of node $i$, denoted by $d_i$, is the cardinality of this set. Likewise, the set of \emph{out-neighbors} of node $i$ is defined as $\mathcal{N}_i^{\mathrm{out}} :=\{j\in\mathcal{V}: (i,j)\in\mathcal{E} \}$ and the {\em out-degree} is denoted by $d_i^{\mathrm{out}} := \abs{\sN_i^{\mathrm{out}}}$. In the rest of the paper, $d_{\max}^{\mathrm{out}}$ is the maximum outgoing degree in the network, that is, $d_{\max}^{\mathrm{out}} := \max_{i\in\mathcal{V}}\{d_i^{\mathrm{out}}\}$. We also denote the minimum in-degree by $d_{\min}^{\mathrm{in}} := \min_{i\in\mathcal{V}}\{d_i\}$. For a digraph $\mathcal{G(\sV,\sE)}$, a nonempty subset $\mathcal{S}\subset\mathcal{V}$ is an \emph{$r$-reachable} set if there exists a node $i \in \mathcal{S}$ such that $\abs{\mathcal{N}_i\setminus \mathcal{S}}\geq r$. That is, the subset $\mathcal{S}$ has at least $r$ in-neighbors in $\mathcal{N}_i\setminus \mathcal{S}$ \citep{leblanc2013resilient,vaidya2012iterative}. This leads to the following concept of  \emph{$r$-robustness} for the digraph $\mathcal{G}(\sV,\sE)$ that characterizes its local redundancy of information flow \citep{leblanc2013resilient,vaidya2012iterative,zhang2015notion}.
\begin{definition}
A digraph $\mathcal{G(V,E)}$ is \emph{$r$-robust} if for every pair of non-empty, disjoint subsets of $\mathcal{V}$, at least one of the subsets is $r$-reachable.
\end{definition}
As in \citep{leblanc2013resilient}, we say that a set $\mathcal{S}\subset \mathcal{V}$ is \emph{$f$-total} if $|\mathcal{S}|\leq f$, $f\in\mathbb{Z}_{\geq 0}$. The set $\mathcal{S}$ is said to be \emph{$f$-totally bounded} if it is an $f$-total set.
\subsection{Stochastic consensus}
\label{sec:app:huang}
In \citep{huang2010automatica,huang2012stochastic} consensus algorithms were considered, which can be written as
\begin{equation}
\label{eq:app:stochastic_approx}
\theta(k+1)=A(k) \theta(k) + H(k) v(k), \quad k\geq 0.
\end{equation}
In the above equation: (i) $\theta(k)=[\theta_1(k) \dots \theta_n(k)]^T\in\mathbb{R}^{n}$ denotes the states of $n$ agents at time $k$; (ii) the initial condition is $\theta_0 = [\theta_{0,1},\ldots,\theta_{0,n}]^T$, with $\E[\lVert\theta_0\rVert^2] < \infty$; (iii) $v(k)\in\mathbb{R}^{m}$ is a noise vector, with $\{v(k), k \ge 0\}$ being a sequence of independent random vectors of zero mean such that $\sup_{k\ge 0} \E[\lVert v(k)\rVert^2] < \infty$. Moreover, $\{A(k),\, k\geq 0\}$ and $\{H(k),\, k\geq 0\}$ are two sequences of random matrices of appropriate dimensions. For each fixed $\omega\in\Omega$, $A(k,\omega)$ is a row-stochastic matrix for all $k\geq 0$. We can now give the following from  \citep{huang2010stochastic,huang2010automatica,huang2012stochastic}.
\begin{definition}
The agents of a network are said to reach \emph{mean square consensus} if $\E\left[ |\theta_i(k)|^2 \right]< \infty$, $\forall k\geq 0$, $i\in\mathcal{V}$, and there exists a random variable $\theta_\infty$ such that $\lim_{k\to\infty} \E\left[ |\theta_i(k)-\theta_\infty|^2 \right]=0$ for all $i\in\mathcal{V}$.
\end{definition}
In the above definition, convergence of $\theta_i(k)$ to $\theta_{\infty}$ is meant as convergence in mean square and in the rest of the paper we simply say that $\theta_i(k)$ converges to $\theta_{\infty}$. As in \citep{huang2012stochastic}, we  define the backward product 
\begin{equation*}
\Psi_{k,s} := A(k-1) \dots A(s), \ \ \mbox{for } k>s\geq 0, \ \ \Psi_{s,s}:=I.
\end{equation*}
Let $\Psi_{k,s}(i,j)$ be the $(i,j)$-th element of $\Psi_{k,s}$. We say that the backward product of the sequence of stochastic matrices $\{A(k),\, k\geq 0\}$ is \emph{ergodic} if $\lim_{k \to \infty} |\Psi_{k,s}(i_1,j)-\Psi_{k,s}(i_2,j)|=0$,
for any given $s\geq 0$, and $i_1$, $i_2$, $j$. 
The following result from \citep[Theorem 3]{huang2012stochastic} is also used in this paper.
\begin{theorem}
\label{thm:app:huang}
Consider (\ref{eq:app:stochastic_approx}) and assume that: 
\begin{description}
\item[(i)] $\{v(k), k\geq 0\}$ is a sequence of independent random vectors of zero mean with bounded variance; 
\item[(ii)] $\sum_{k=0}^\infty \E\left[ \lVert H(k)\rVert_F^2 \right] \E\left[ v(k)^T v(k) \right] < \infty$;
\item[(iii)] $\{A(k),\, k\geq 0\}$ has ergodic backward product with probability $1$.
\end{description}
Then, for any initial state $\theta_0$ with bounded variance, system \eqref{eq:app:stochastic_approx} achieves mean square consensus.
\end{theorem}
\section{Problem Formulation}
\label{sec:problem_formulation}
We now formalize the problem statement and, in order to do so, we first introduce the notions of adversary agent, resilient consensus and differential privacy. To this aim, consider a multi-agent system of $n$ agents evolving on a digraph $\mathcal{G}(\mathcal{V},\mathcal{E})$. The internal state of the $i$-th agent at time $k\in\mathbb{Z}_{\geq 0}$ is denoted by $\theta_i (k) \in \R$ and by $\theta_{i,0} := \theta_i(0)$ its initial condition. The message sent at time $k$ from agent $i$ is $x_i(k)\in\mathbb{R}$, while the messages received by agent $i$ at time $k$ from its in-neighbors are denoted by $\{x_j(k), j \in\sN_i\}$. The internal state of agent $i\in\mathcal{V}$ is updated according to some update rule of the form
\begin{equation}
\label{eq:general_update_rule}
\theta_i(k+1)=g_i\left( k, \theta_i(k), \{ x_j(k), j \in\sN_i \} \right), \ k\in\mathbb{Z}_{\geq 0},
\end{equation}
while the message sent by the $i$-th agent at time $k$ is generated as
\begin{equation}
\label{eq:general_message}
x_i(k)=h\left(\theta_i(k),\eta_i(k)\right), \ \ k\in\mathbb{Z}_{\geq 0},
\end{equation}
with $\eta_i(k)\in\mathbb{R}$ being some noise injected by agent $i$ at time $k$ with some arbitrary distribution. We remark that the update rule $g_i(\cdot)$ may be different from agent to agent, while the function $h(\cdot)$ is the same for all the agents, i.e. agents use the same communication policy.
\subsection{Adversary model and resilient consensus}
\label{sec:resilient_consensus}
Suppose that the network node set $\mathcal{V}$ is partitioned into a set of normal or \emph{non-faulty} nodes, evolving according to \eqref{eq:general_update_rule} and \eqref{eq:general_message}, and a set of adversary or \emph{faulty} nodes $\mathcal{F}\subset\mathcal{V}$. Moreover, $|\mathcal{F}|=\phi\leq f$, i.e. the set of faulty nodes is $f$-totally bounded (only the upper bound $f$ is known and not the actual $\phi$). We use the following classification from \citep{leblanc2013resilient}. 
\begin{definition}
An agent $i\in\mathcal{F}$ is said to be \emph{Byzantine} if it does not send the same value to all of its neighbors at some time-step, or if it does not follow the same update rule as the non-faulty agents. An agent $i\in\mathcal{F}$ is said to be \emph{malicious} if it sends the same value to all of its neighbors, but it does not follow the same update rule as the non-faulty agents.
\end{definition}
Both malicious and Byzantine agents are allowed to update their states \emph{arbitrarily} and can collude among them. Furthermore, the faulty agents are often assumed to be \emph{omniscient}, i.e. they know the network topology, the dynamics of the other agents and their initial conditions. Note that all malicious agents are Byzantine but not vice versa. Therefore, Byzantine agents are more general and in this paper we use the terminology {\em faulty agent} to denote both Byzantine and malicious agents. 

Each faulty agent $i\in\mathcal{F}$ can send a different, arbitrary, attack signal $\widehat{x}_i^j(k)$ to each of its out-neighbors, $j\in\mathcal{N}_i^{\mathrm{out}}$.
The attack signals $\widehat{x}_i^j$ are injected by the faulty agents into the network to steer the multi-agent system towards some  malicious consensus state $\widehat{\theta}_\infty$. In applications, this malicious state $\widehat{\theta}_\infty$ may either be dangerous for the integrity of the multi-agent system or it may be defined to corrupt the computation carried out by the network. 
Given that faulty agents are omniscient (worst case situation), the signal $\widehat{x}_i^j(k)$ depends on the malicious consensus state $\widehat{\theta}_\infty$ and on the present and past internal state history of non-faulty nodes $\mathcal{V\setminus F}$. That is,
\begin{equation}
\label{eq:general_faulty_message}
\widehat{x}_i^j(k)=\widehat{h}_i^j\big(\widehat{\theta}_\infty; \theta(0),\dots,\theta(k)\big), \ \ k\in\mathbb{Z}_{\geq 0},
\end{equation}
where $\theta(k) = [\theta_1(k),\ldots,\theta_{n-\phi}(k)]^T$ is the stack vector of internal state of non-faulty nodes at time $k$. In the context of this paper, the non-faulty agents neither are aware of the attack law in \eqref{eq:general_faulty_message} nor are interested to detect and isolate the faulty agents. In what follows, we denote with the {\em hat} all the quantities related to faulty agents.

We are now ready to give the following definition, adapted from \citep{leblanc2013resilient,vaidya2012iterative,huang2009coordination}.
\begin{definition}\label{def:correctness}
The non-faulty agents are said to achieve \emph{resilient asymptotic consensus} if for any choice of their initial values the following two conditions are verified:
\begin{itemize}
\item
\emph{Validity}:
$
\min_{j\in\mathcal{V}\setminus\mathcal{F}} \mathbb{E}\big[ \theta_j(k) \big] \leq \mathbb{E}\big[ \theta_i(k+1) \big] \leq \max_{j\in\mathcal{V}\setminus\mathcal{F}} \mathbb{E}\big[ \theta_j(k) \big],
$
for all $k\geq 0$ and $i\in\mathcal{V\setminus F}$;
\item
\emph{Convergence}: 
$
\lim_{k\to\infty} \E\big[|\theta_i(k)-\theta_j(k)|^2 \big] =0,
$
for all $i,j\in\mathcal{V\setminus F}$, $i\neq j$. 
\end{itemize}
\end{definition}
Definition \ref{def:correctness} implies that non-faulty agents achieve mean square consensus to a value that is between the smallest and largest values of their initial conditions. A consensus algorithm that guarantees both the validity and the convergence conditions is said to be \emph{correct}, see \citep{leblanc2013resilient,vaidya2012iterative,huang2009coordination}.
\subsection{Differential privacy}
\label{sec:differential_privacy}
We introduce differential privacy by closely following \citep{nozari2017differentially}. Here, we let $\theta(k)$, $x(k)$ and $\eta(k)$ be the stack vectors of all states/attack signals, messages sent and generated noises at time $k$ in the whole network, respectively. These vectors are in $\mathbb{R}^m$, with $m=(n-\phi)+\sum_{i\in\mathcal{F}} d_i^{\mathrm{out}}$.
We denote by $\bm{\eta}=\{\eta(k)\}_{k=0}^\infty$ the sequence of vector noise on the total sample space $\Omega=(\mathbb{R}^m)^{\mathbb{N}}$. Given an initial state $\theta_0$, the sequences of random vector variables $\mathbf{x}=\{x(k)\}_{k=0}^\infty$ and $\bm{\theta}=\{\theta(k)\}_{k=0}^\infty$ are uniquely determined by $\bm{\eta}$. Therefore, the function $X_{\theta_0}: (\mathbb{R}^m)^{\mathbb{N}} \rightarrow (\mathbb{R}^m)^{\mathbb{N}}$ such that $X_{\theta_0}(\bm{\eta})=\mathbf{x}$ is well defined and represents the \emph{observation} corresponding to the \emph{execution} $(\theta_0,\bm{\eta})$.
\begin{definition}\label{def:diff_privacy}
Given $\delta\in\mathbb{R}_{\geq 0}$, the initial states of the non-faulty agents of the network $\theta_0^{(1)}$ and $\theta_0^{(2)}$ are \emph{$\delta$-adjacent} if, for some $i_0\in\mathcal{V}\setminus \mathcal{F}$, $|\theta_{0,i}^{(2)}-\theta_{0,i}^{(1)}|\leq \delta$ if $i=i_0$ and $|\theta_{0,i}^{(2)}-\theta_{0,i}^{(1)}| = 0$ if $i\neq i_0$, for $i\in\mathcal{V}\setminus \mathcal{F}$. Given $\delta,\varepsilon\in\mathbb{R}_{\geq 0}$, the network dynamics is \emph{$\varepsilon$-differentially private} if, for any pair $\theta_0^{(1)}$ and $\theta_0^{(2)}$ of $\delta$-adjacent initial states and any set $\mathcal{O}\in\mathcal{B}\left( (\mathbb{R}^{n-\phi})^{\mathbb{N}} \right)$,
\begin{equation}
\label{eq:def_diff_priv}
\sP\{ \bm{\eta}\in\Omega | X_{\theta_0^{(1)}}(\bm{\eta})\in \mathcal{O} \} \leq e^{\varepsilon} \sP\{ \bm{\eta}\in\Omega | X_{\theta_0^{(2)}}(\bm{\eta})\in \mathcal{O} \}.
\end{equation}
\end{definition}
We also give a definition of accuracy, which is adapted from \citep{nozari2017differentially}.
\begin{definition}
For $p\in[0,1]$ and $r\in\mathbb{R}_{\geq 0}$, the network dynamics is \emph{$(p,r)$-accurate} if, for any initial condition, the state of each non-faulty agent converges, in any sense, as $k\to\infty$ to a random variable $\theta_\infty$, with $\E\left[ \theta_\infty \right]<\infty$, and $\sP\{|\theta_\infty - \E[\theta_\infty]|\leq r\} \geq 1-p$.
\end{definition}
\subsection{Problem statement}
\label{sec:problem_statement}
Given the set-up described above, consider a multi-agent system and assume that a $f$-totally bounded set of faulty agents exists, $\mathcal{F}\subset\mathcal{V}$. Our goal in this paper is to design a distributed consensus algorithm ensuring that: (i) resilient asymptotic consensus is achieved by the non-faulty agents with $(p,r)$-accuracy. That is, the algorithm allows the non-faulty agents to achieve consensus despite the presence of faulty agents trying to steer the multi-agent system toward some malicious consensus state $\widehat{\theta}_\infty$; (ii) $\varepsilon$-differential privacy of the initial conditions of non-faulty agents is guaranteed when faulty agents behave arbitrarily within the communication policy adopted in the system. The algorithm presented in this paper consists of: (i) an update law of the form \eqref{eq:general_update_rule}; (ii) an inter-agent message generator of the form \eqref{eq:general_message}; (iii) the distribution of the noise processes $\eta_i(k)$.
\section{The DP-MSR algorithm}
\label{sec:algorithm}
We propose a modification of the original MSR algorithm, which we term as Differentially Private MSR algorithm (DP-MSR). The key steps of the DP-MSR algorithm are summarized as pseudo-code in Algorithm \ref{alg:DP-MSR}. An agent that follows the DP-MSR: (i) transmits to its neighbors a message that is {\em corrupted} with noise (rather than its internal state as in the MSR). Note that the {\em Transmit phase} in Algorithm \ref{alg:DP-MSR} defines the communication policy of the network; (ii) discards, as in the MSR, some of the (noisy) messages from the neighbors before updating its internal state. In Section \ref{sec:properties} we characterize correctness, accuracy and differential privacy of the DP-MSR algorithm. In particular, we show that these properties are related to the topology of the underlying graph $\mathcal{G}$.
\begin{algorithm}[!t]
\begin{algorithmic}
\caption{DP-MSR}\label{alg:DP-MSR}
{\small
	\State {\bf Variables:}
	\State  $\theta_i(k)$ internal state of agent $i$ at time $k$
	\State $\eta_i(k)$ independent zero-mean noise with Laplace distribution, i.e. $\eta_i(k)\sim \mathrm{Lap}(b(k))$, with: $b(k)=c\, q^k$, and $c>0$, $\frac{1}{2}< q <1$ being two design parameters
	\State $f$ design parameter 
	\vspace{0.2cm}
	\State {\bf Initialization:}
	\State Set $\theta_i(0)\gets\theta_{0,i}$ and $a_i\gets 1/\left(|\mathcal{N}_i|-2f+1\right)$
	\vspace{0.2cm}
	\While {True}
	\State {\bf Transmit phase}
	\State 	Send to each out-neighbor $x_i(k) \gets \theta_i(k)+\eta_i(k)$
	\State {\bf Receive phase}
	\State Get $x_j(k)$ from each in-neighbor
	\State Store the values in the vector $r_i(k)$ of size $\abs{\sN_i}$
	\State {\bf Update phase}	
	\State Remove from $r_i(k)$ the largest and smallest $f$ values
	\State Compute:
	\begin{equation*}
\theta_i(k+1) \gets a_i\theta_i(k)+\sum_{j\in \mathcal{N}_i^\ast (k)} a_i x_j(k),
\end{equation*} 
where $\mathcal{N}_i^\ast (k)$ is the set of in-neighbors that have not been removed from $r_i(k)$
	\EndWhile
}
\end{algorithmic}
\end{algorithm}

Following Algorithm \ref{alg:DP-MSR}, each agent $i$ executes at every time step $k$ the following three actions:
\begin{enumerate}
\item
\emph{Transmit phase}: agent $i$ transmits to all its out-neighbors the message $x_i$, generated as
\begin{equation}
\label{eq:adding_noise}
x_i(k) = \theta_i(k)+\eta_i(k),
\end{equation}
where $\eta_i(k)$ is a zero-mean noise with Laplacian distribution $\eta_i(k)\sim \mathrm{Lap}(b(k))$ and $b(k)=c\, q^k$, $c>0$, $\frac{1}{2}< q < 1$;
\item
\emph{Receive phase}: agent $i$ receives the messages $\{ x_j(k), j \in \sN_i\}$ from all its in-neighbors. These values are used to create the vector $r_i(k)$ with dimension $|\mathcal{N}_i(k)|$;
\item
\emph{Update phase}: agent $i$ stores the values contained in $r_i(k)$ and removes from such a vector the smallest $f$ values and the largest $f$ values (breaking ties arbitrarily). We denote by $\mathcal{N}_i^\ast (k)$ the set of in-neighbors of the $i$-th agent whose values have not been removed. Clearly, $\abs{\sN_i^\ast (k)} = |\mathcal{N}_i(k)|-2f$. Then, agent $i$ updates its internal state as
\begin{equation}
\label{eq:update_rule}
\theta_i(k+1) = a_i\theta_i(k)+\sum_{j\in \mathcal{N}_i^\ast (k)} a_i x_j(k),
\end{equation}
with $a_i = \frac{1}{\abs{\sN_i^\ast (k)}+1} = \frac{1}{|\mathcal{N}_i|-2f+1}$ (that is, the agent  takes the average of its internal state $\theta_i(k)$ and the remaining messages in $r_i(k)$).
Note that the $a_i$'s only depend on $f$ and on the topology of the graph $\mathcal{G}(\mathcal{V},\mathcal{E})$.
\end{enumerate}
\section{Properties of the DP-MSR algorithm}
\label{sec:properties}
We now investigate correctness, accuracy and differential privacy of Algorithm \ref{alg:DP-MSR}.
\subsection{Correctness}
\label{sec:correctness}
We show that if the network topology is sufficiently robust, then Algorithm \ref{alg:DP-MSR} is correct according to Definition \ref{def:correctness}. This is done by proving that: (i) each non-faulty agent $i\in\mathcal{V}\setminus\mathcal{F}$ (described by the stochastic process $\theta_i(k)$) converges in mean square to some common random variable $\theta_\infty$, despite the presence of up to $f$ faulty agents in the network (convergence condition); (ii) for each non-faulty agent $i\in\mathcal{V}\setminus\mathcal{F}$, $\E[\theta_i(k+1)]$ is the convex combination of $\left\{ \E[\theta_j(k)] \right\}$, ${j\in\mathcal{V}\setminus\mathcal{F}}$, for $k\in\mathbb{Z}_{\geq 0}$ (validity condition).
\begin{lemma}
\label{app:lemma_vaidya}
Consider the multi-agent system \eqref{eq:adding_noise}-\eqref{eq:update_rule} and without loss of generality suppose that the first $n-\phi$ agents are non-faulty and let $\theta(k) = [\theta_1(k),\ldots,\theta_{n-\phi}(k)]^T$. Assume that the network topology is at least $(2f + 1)$-robust, then the dynamics of the non-faulty agents can be expressed as
\begin{equation}
\label{eq:ffree_node_update}
\theta_i(k+1)= M_i(k) \theta(k) + \sum_{j\in \mathcal{N}_i^\ast (k)} a_i \eta_j(k),
\end{equation}
with $i\in\{1,\dots, n-\phi\}$ and  $M_i(k)$ being a row-stochastic vector of size $(n-\phi)$ having the following properties: (1) the $i$-th element of $M_i(k)$, say $M_{ii}(k)$, is such that $M_{ii}(k) = a_i$; (2) the $j$-th element of $M_i(k)$, i.e. $M_{ij}(k)$ is non-zero only if $(j,i) \in\sE$; (3) $\forall k$, at least $|\mathcal{N}_i\cap(\mathcal{V}\setminus\mathcal{F})|-f+1$ elements of $M_i(k)$ are lower bounded by some positive constant.
\end{lemma}
\begin{pf*}{Proof.}
This lemma can be proved by using arguments similar to those presented in \citep{vaidya2012matrix,su2016reaching}, where an analogous result was proved in the case where there is no noise in the network (i.e. $\eta_i = 0$, $\forall i = 1,\ldots,n$). By combining \eqref{eq:adding_noise} and \eqref{eq:update_rule} the closed loop dynamics of non-faulty agents can be written as
\begin{equation}
\label{eqn:closed_loop}
\begin{split}
\theta_i(k+1)&= a_i\theta_i(k)+\sum_{j\in \mathcal{N}_i^\ast (k)} a_i (\theta_j(k)+\eta_j(k))\\
&= \sum_{j\in \{i\}\cup \mathcal{N}_i^\ast (k)} a_i \theta_j(k) + \sum_{j\in \mathcal{N}_i^\ast (k)} a_i \eta_j(k).
\end{split}
\end{equation}
Note that the set $\mathcal{N}_i^\ast(k)$ in the above dynamics may contain values coming from faulty agents. Indeed, faulty agents might generate values between the $f$ largest and the $f$ smallest values received by a non-faulty agent and therefore those values would not be eliminated by the algorithm. In this case, as shown in details in \citep{vaidya2012matrix}, under the hypothesis that the network topology is $(2f+1)$-robust, it is always possible to express the value $\theta_j(k)$ from any faulty agent in $\mathcal{N}_i^\ast(k)$ as the convex combination of those of the non-faulty agents in $\mathcal{N}_i \setminus \mathcal{N}_i^\ast(k)$, that were removed ``accidentally'' by the algorithm. \hfill$\square$
\end{pf*}
Lemma \ref{app:lemma_vaidya} states that the time evolution of the non-faulty agent can be conveniently expressed as in (\ref{eq:ffree_node_update}) and, moreover, the noise-free term $M_i(k)\theta(k)$ depends only on the non-faulty agents. 
\begin{proposition}
\label{prop:correctness}
\label{thm:mean_convergence}
Consider the multi-agent system \eqref{eq:adding_noise}-\eqref{eq:update_rule}. Assume that the network topology is at least $(2f+1)$-robust. Then, the DP-MSR algorithm is correct and the closed loop dynamics \eqref{eqn:closed_loop} reaches mean square consensus, i.e. there exists a random variable $\theta_\infty$ such that $\lim_{k\to\infty} \E\left[ |\theta_i(k)-\theta_\infty|^2 \right]=0$ for all $i\in\mathcal{V}\setminus\mathcal{F}$.
\end{proposition}
\begin{pf*}{Proof.}
Without loss of generality, the attack signal \eqref{eq:general_faulty_message} can be modeled as the second order random process 
\begin{equation*}
\widehat{x}_i^j(k)= \widehat{\theta}_i^j(k)+\widehat{\eta}_i^j(k),
\end{equation*}
where $\widehat{\theta}_i^j(k):\mathbb{Z}_{\geq 0} \rightarrow \mathbb{R}$ is some arbitrary deterministic signal, and $\widehat{\eta}_i^j(k)$ is a zero-mean white noise with bounded variance $\widehat{\sigma}_{i,j}^2(k)$, for all $k\geq 0$, and such that $\sum_{k=0}^{\infty} \widehat{\sigma}_{i,j}^2(k) <\infty$.
\\
Now, from Lemma \ref{app:lemma_vaidya} the evolution of a non-faulty agent can be expressed as in \eqref{eq:ffree_node_update}. 
Therefore, by taking the expected value we get $\E\left[ \theta_i(k+1) \right]= M_i(k) \E\left[ \theta(k) \right]$, $k\geq 0.$
Since $M_i$ is a stochastic vector, then the validity condition as in Definition \ref{def:correctness} is satisfied for all $k\geq 0$.\\
We can prove convergence by means of Theorem \ref{thm:app:huang} after noticing that \eqref{eq:ffree_node_update} can be written in compact form as
\begin{equation}
\label{eq:ffree_update}
\theta(k+1)=M(k)\theta(k)+\breve{\eta}(k),
\end{equation}
where $M(k)$ is a $(n-\phi)\times (n-\phi)$ row-stochastic matrix and
\begin{equation}
\label{eq:noise_components}
\breve{\eta}_i(k)=\sum_{j \in \mathcal{N}_i^\ast (k)\setminus \mathcal{Q}_i(k)} a_i \eta_j(k) + \sum_{j \in \mathcal{Q}_i(k)} a_i \widehat{\eta}_j^i(k),
\end{equation}
with $\mathcal{Q}_i(k)=\mathcal{N}_i^\ast(k) \cap \mathcal{F}$ being the set of faulty agents in $\mathcal{N}_i^\ast (k)$ at time $k$. 
Indeed, observe that the dynamics \eqref{eq:ffree_update} has the same form as \eqref{eq:app:stochastic_approx}, with $A(k)=M(k)$, $H(k)=I_{n-\phi}$, $v(k) = \breve\eta(k)$ and that hypothesis (i) of such theorem is satisfied by construction (recall indeed that $\breve{\eta}(k)$ is a sequence of independent random vectors with zero-mean and bounded covariance).
We now show that hypotheses (ii) and (iii) of Theorem \ref{thm:app:huang} are also satisfied by \eqref{eq:ffree_update}. 
To show that (ii) holds, we note that, in this case, $ \sum_{k=0}^\infty \E\left[ \mathrm{tr} \left( H(k)^T H(k) \right) \right] \E\left[ v(k)^T v(k) \right]  = (n-\phi)\, \sum_{k=0}^\infty \E\left[ \breve{\eta}^T(k)\breve{\eta}(k) \right]$ and hence
\begin{equation*}
\label{eq:totalnoise_variance}
\begin{split}
\sum_{k=0}^\infty & \E\left[ \breve{\eta}^T(k)\breve{\eta}(k) \right] \\
& \leq \sum_{k=0}^\infty \sum_{i=1}^{n-\phi} a_i^2 \left( \left|\mathcal{N}_i^\ast(k)\setminus \mathcal{Q}_i(k)\right| 2b^2(k) + \left| \mathcal{Q}_i(k)\right|  \widehat{\sigma}_{j,i}^2(k)    \right)\\
& \leq \sum_{k=0}^\infty \sum_{i=1}^{n-\phi} a_i^2 \left|\mathcal{N}_i^\ast(k)\right| \max\{2b^2(k),\widehat{\sigma}_{j,i}^2(k)\} \\
& \leq (n-f)^2 \sum_{k=0}^\infty \max\{2b^2(k),\widehat{\sigma}_{j,i}^2(k)\} <\infty,
\end{split}
\end{equation*}
where we used the fact that $0<a_i\leq 1$, that the variance of the noise signals goes to zero, and that, for each agent $i$, the quantity $|\mathcal{N}_i^\ast(k)|=|\mathcal{N}_i|-2f$ is constant for all $k\geq 0$ and such that 
$
|\mathcal{N}_i^\ast(k)|\leq  (n-1)-2f \leq n-f.
$
Now, hypothesis (iii) of Theorem \ref{thm:app:huang} can be shown by noting that, since the network is $(2f+1)$-robust, then from  \citep[Theorem 1]{vaidya2012matrix} we can immediately conclude that the sequence $\{M(k),\, k\geq 0\}$ has ergodic backward product. Therefore, Theorem \ref{thm:app:huang} holds and hence we can conclude that there exists a random variable $\theta_\infty$ such that $\lim_{k\to\infty} \E\left[ |\theta_i(k)-\theta_\infty|^2 \right]=0$, for all $i\in\mathcal{V}\setminus\mathcal{F}$. This proves the result. \hfill$\square$
\end{pf*}
\subsection{Accuracy}
Next we analyze the statistical properties of $\theta_\infty$, together with the accuracy of the algorithm. 
\begin{proposition}\label{prop:accuracy}
Consider the multi-agent system \eqref{eq:adding_noise}-\eqref{eq:update_rule}. Assume that the network topology is at least $(3f+1)$-robust, then the random variable $\theta_\infty$ towards which the state of non-faulty agents, $\theta_i(k)$, converges as $k\to\infty$ is such that $\E\left[ \theta_\infty \right] = w^T\theta_0$, for some vector $w\in\mathbb{R}^n$ having nonnegative elements with $\mathbf{1}^Tw=1$, and 
\begin{equation}
\label{eq:variance_bounds}
\frac{ 2\, c^2 \min_i a_i^2  }{n(1-q^2)} \leq \mathrm{var}\left( \theta_\infty \right) \leq \frac{c^2\, (n-f)}{2(1-q^2)} .
\end{equation}
\end{proposition}
\begin{pf*}{Proof.}
Proposition \ref{thm:mean_convergence} implies that $\lim_{k\to\infty} \E\big[ \Vert \theta(k)-\theta_\infty\mathbf{1}\Vert^2 \big]=0$,
which in turns implies that $\lim_{k\to\infty} \E\left[ \theta(k) \right] = \E\left[\theta_\infty \mathbf{1} \right]$. 
Denoting with $\Psi_{k,s}$ the backward product of the sequence $\{M(k),\,k\geq 0\}$, 
from equation \eqref{eq:ffree_update} we get
\begin{equation*}
\theta(k)=\Psi_{k,0}\,\theta_0 +  \sum_{h=0}^{k-1} \Psi_{k,h+1} \breve{\eta}(h), \ \ k\geq 0.
\end{equation*}
Now, from the fact that $\E\left[ \breve{\eta}(k) \right]=0$, $\forall k\geq 0$, and since the sequence $\{M(k),\, k\geq 0\}$ has ergodic backward product, it follows that 
$\lim_{k\to\infty} \E\left[ \theta(k)\right] = \lim_{k\to\infty} \Psi_{k,0}\,\theta_0 = \mathbf{1} w^T \theta_0$,
for some $w\in\mathbb{R}^n$ such that $\mathbf{1}^Tw=1$ \cite[Lemma 3.7]{ren2005consensus}. Hence, we have that $\E\left[ \theta_\infty \right]= w^T\theta_0$. Now, since
\begin{equation*}
\begin{split}
\theta(k)^T \theta(k) & = \theta_0^T \Psi_{k,0}^T \Psi_{k,0} \theta_0
 + 2 \sum_{h=0}^{k-1} \theta_0^T \Psi_{k,0}^T \Psi_{k,h+1} \breve{\eta}(h)\\
& +  \sum_{h=0}^{k-1} \breve{\eta}^T(h)\Psi_{k,h+1}^T \Psi_{k,h+1} \breve{\eta}(h)
\end{split}
\end{equation*}
and 
$$
\E\left[\sum_{h=0}^{k-1} \theta_0^T \Psi_{k,0}^T \Psi_{k,h+1} \breve{\eta}(h) \right]= \E\left[ \sum_{h=0}^{k-1} \zeta_h^T\breve{\eta}(h) \right]=0,
$$
for some vectors $\zeta_h\in\mathbb{R}^n$, we have that
\begin{equation*}
\begin{split}
\mathrm{var}(\theta(k))& = \E\left[ \theta^T(k)\theta(k) \right] - \E\left[ \theta(k) \right]^T \E\left[ \theta(k) \right] \\
& = \sum_{h=0}^{k-1} \E\left[ \breve{\eta}^T(h)\Psi_{k,h+1}^T \Psi_{k,h+1} \breve{\eta}(h) \right]\\
& = \sum_{h=0}^{k-1} \sum_{i=1}^{n-\phi} \mathrm{var}\left( \breve{\eta}_i(h)  \right) \left( \Psi_{k,h+1}^T \Psi_{k,h+1} \right)_{ii},
\end{split}
\end{equation*}
having taken into account that $\breve{\eta}_i$ are independent random variables. Now, from Proposition \ref{thm:mean_convergence} it also follows that $\lim_{k\to\infty} \mathrm{var}(\theta_i(k))=\mathrm{var}(\theta_\infty)$ and hence we have 
\begin{equation}
\label{eq:final_variance}
\begin{split}
\mathrm{var}(\theta_\infty) & = \frac{1}{n-\phi} \lim_{k\to\infty} \mathrm{var}(\theta(k))\\
& = \frac{1}{n-\phi}\, \lim_{k\to\infty} \sum_{h=0}^{k-1} \sum_{i=1}^{n-\phi} \mathrm{var}\left( \breve{\eta}_i(h)  \right) \left( \Psi_{k,h+1}^T \Psi_{k,h+1} \right)_{ii}.
\end{split}
\end{equation}
Based on this, we now give upper and lower bounds on $\mathrm{var}(\theta_\infty)$.\\
\emph{Upper bound: } Consider the worst-case scenario where every faulty agent injects noise $\widehat{\eta}_i^j(k)$. Without loss of generality, we can assume that 
$$
\widehat{\sigma}_{i,j}^2(k) \leq 2b^2(k) = 2{c}^2\,{q}^{2k},\ \ k\in\mathbb{Z}_{\geq 0},
$$
i.e. $\widehat{\sigma}_{i,j}^2(k)$ decays at least as the variance of the non-faulty agents in \eqref{eq:adding_noise} and any difference from this is taken into account in $\widehat{\theta}_j^i(k)$.
From the hypothesis on the robustness of the network, we have that $|\mathcal{N}_i|\geq 3f+1$, and so
$a_i=\frac{1}{|\mathcal{N}_i|-2f+1}\leq \frac{1}{f+2}< \frac{1}{2}$,
for every node $i\in\mathcal{V}\setminus\mathcal{F}$.
Hence, from \eqref{eq:noise_components} and recalling that $|\mathcal{N}_i^\ast(k)|\leq n-f$, we have that $\mathrm{var}\left( \breve{\eta}_i(h)  \right) \leq (n-f)a_i^2 \widehat{\sigma}_{j,i}^2(h)\leq \frac{1}{2}(n-f)c^2\, q^{2h}$.
Therefore, from \eqref{eq:final_variance} we get
\begin{equation*}
\begin{split}
\mathrm{var}(\theta_\infty) & \leq \frac{n-f}{2(n-\phi)} \lim_{k\to\infty} \sum_{h=0}^{k-1} \sum_{i=1}^{n-\phi} c^2\, q^{2h} \left( \Psi_{k,h+1}^T \Psi_{k,h+1} \right)_{ii}\\
& \leq \frac{n-f}{2(n-\phi)}  \lim_{k\to\infty} \sum_{h=0}^{k-1}  c^2\, q^{2h} \, \mathrm{tr}\left( \Psi_{k,h+1}^T \Psi_{k,h+1} \right)\\
& \leq \frac{n-f}{2} \sum_{h=0}^\infty c^2\, q^{2h} =  \frac{c^2\, (n-f)}{2(1-q^2)},
\end{split}
\end{equation*}
where we used the fact that, for any $m\times m$ row-stochastic matrix $\Psi$, it holds that $\mathrm{tr}(\Psi^T\Psi)=\mathrm{tr}(\Psi \Psi^T)\leq m$.\\
\emph{Lower bound: } In the case that the faulty agents do not add any noise to their signals, i.e. $\widehat{\eta}_j^i(k)=0$, from \eqref{eq:noise_components} we have that $\mathrm{var}\left( \breve{\eta}_i(h)  \right) = \left| \mathcal{N}_i^\ast(h)\setminus \mathcal{Q}_i(h)  \right| a_i^2 2b^2(h)$. Again, from the hypothesis on the robustness of the network we have that $\left| \mathcal{N}_i^\ast(h)\setminus \mathcal{Q}_i(h)  \right| = |\mathcal{N}_i|-2f-|\mathcal{Q}_i(h)| \geq 3f+1-2f-\phi\geq 1$ for all $i$ and $h\geq 0$. Hence, from \eqref{eq:final_variance} we get
\begin{equation*}
\begin{split}
\mathrm{var}(\theta_\infty) & \geq \frac{1}{n-\phi}  \lim_{k\to\infty} \sum_{h=0}^{k-1} \sum_{i=1}^{n-\phi} a_i^2 2{b}^2(h) \left( \Psi_{k,h+1}^T \Psi_{k,h+1} \right)_{ii}\\
& \geq \frac{1}{n} \lim_{k\to\infty} \sum_{h=0}^{k-1} \left[ \min_i (a_i^2) 2c^2q^{2h} \right]  \mathrm{tr}\left( \Psi_{k,h+1}^T \Psi_{k,h+1} \right)\\
& \geq \frac{ 2\, c^2 \min_i a_i^2  }{n} \sum_{h=0}^\infty q^{2h} = \frac{ 2\, c^2 \min_i a_i^2  }{n(1-q^2)},
\end{split}
\end{equation*}
where we used the fact that, for any row-stochastic matrix $\Psi$, it holds that $\mathrm{tr}(\Psi^T\Psi)=\lVert \Psi\rVert^2_F\geq 1$. \hfill$\square$
\end{pf*}
\begin{remark}
\label{rem:3}
Note that $(3f+1)$-robustness is required to obtain a nonzero lower bound in (\ref{eq:variance_bounds}) independently on the network topology (the upper bound, instead, only requires $(2f+1)$-robustness). The fact that the network is $(3f+1)$-robust guarantees that, after Algorithm \ref{alg:DP-MSR} removes $2f$ messages coming from in-neighbors, at least one of the remaining $f+1$ messages comes from a noisy non-faulty agent in $\sN_i^{\ast}$. 
On the other hand, even a slight relaxation of this condition, say the network is $3f$-robust, would not allow to get a nonzero lower bound on $\mathrm{var}(\theta_\infty)$ for \emph{every} network topology. Indeed, for any $3f$-robust network we know that $|\mathcal{N}_i|\geq 3f$, for all $i$, hence $|\mathcal{N}_i^\ast(h)\setminus \mathcal{Q}_i(h)|=|\mathcal{N}_i|-2f-|\mathcal{Q}_i(h)| \geq 3f-2f-\phi \geq f-f =0$. 
Therefore, if the messages received by some non-faulty agent do not contain any noise, i.e. $\widehat{\eta}_j^i(h)=0$, $\forall i,j$, $h\geq0$, then from \eqref{eq:noise_components} it follows that for some $i$ and $h$ it may occur that $\mathrm{var}(\breve{\eta}_i(h))=0$. 
In other words, there exists a $3f$-robust network such that the lower bound on $\mathrm{var}(\theta_\infty)$ is zero.
\end{remark}
\begin{proposition}
\label{prop:accuracy_2}
Under same assumptions of Proposition \ref{prop:accuracy}, Algorithm \ref{alg:DP-MSR} is  $\left(p, c \sqrt{\frac{(n-f)}{2p(1-{q}^2)}} \right)$-accurate for any ${p\in(0,1)}$.
\end{proposition}
\begin{pf*}{Proof.}
From Chebyshev's inequality it follows that
\begin{equation*}
\sP  \{ |\theta_\infty - w^T\theta_0|\leq r \} \geq 1- \frac{\mathrm{var}(\theta_\infty)}{r^2} \geq 1-\frac{\overline{\mathrm{var}}(\theta_\infty)}{r^2},
\end{equation*}
where we denoted by $\overline{\mathrm{var}}(\theta_\infty)$ the upper bound in \eqref{eq:variance_bounds}. Now, by picking $p=\overline{\mathrm{var}}(\theta_\infty)/r^2$ it follows that
$r=\sqrt{\frac{\overline{\mathrm{var}}(\theta_\infty)}{p}}=c \sqrt{\frac{(n-f)}{2p(1-{q}^2)}}$.
\hfill$\square$
\end{pf*}
As in \citep{nozari2017differentially}, it follows that the ideal case of $(0,0)$-accuracy cannot be obtained because the radius $r$ is a decreasing function of $p$.
\subsection{Differential privacy}
\label{sec:diff_privacy_analysis}
We now investigate the $\varepsilon$-differential privacy properties of the algorithm. To this aim, we consider two important special cases: (i) absence of faulty agents, i.e. $\phi=0$;  (ii) presence of $\phi \ne 0$ faulty agents that, similarly to \citep{sundaram2011distributed}, behave arbitrarily within the communication policy of the network, i.e. $\widehat{\eta}_i^j(k)\sim \mathrm{Lap}(\widehat{b}(k))$, with $\widehat{b}(k)=c\,q^k$. A key difference between the model considered here and the one in \citep{sundaram2011distributed} is that in the latter it is assumed that faulty agents send the same value to their neighbors. 
In our case, omniscient faulty agents have the capability to adapt their attack signals to the initial conditions of non-faulty agents to drive the multi-agent system towards some malicious consensus value $\widehat{\theta}_\infty$. 
Specifically, let $\theta_0^{(1)}$ and $\theta_0^{(2)}$ be two $\delta$-adjacent initial conditions of non-faulty agents and let $\widehat{\theta}_i^{j,(1)}(k)$ and $\widehat{\theta}_i^{j,(2)}(k)$ be the corresponding input attack signals injected by the $i$-th faulty agent to steer the multi-agent system towards the \emph{same} $\widehat{\theta}_\infty$. Then, without loss of generality, the two input signals are such that
\begin{equation}
\label{eq:attack_bound}
\widehat{\theta}_i^{j,(2)}(k) = \widehat{\theta}_i^{j,(1)}(k) + \widehat{\delta}_i^j(k),
\end{equation}
where $\widehat{\delta}_i^j(k)$ converges to zero \emph{faster} than the noise injected by the agents, that is $\lvert \widehat{\delta}_i^j(k)\rvert \leq \bar\delta\, \lambda^k$, with $\bar{\delta}\geq 1$, $0<\lambda<q$.
\begin{proposition}
\label{prop:privacy}
Consider the multi-agent system \eqref{eq:adding_noise}-\eqref{eq:update_rule} and let $\delta$ be the adjacency bound of the initial conditions of non-faulty agents. The following statements hold: 
\begin{enumerate}
\item[(i)]
if $\phi = 0$, then $\bar{\varepsilon}$-differential privacy of the agents is guaranteed, where 
$\bar{\varepsilon}=\delta \frac{2q}{c\,(2q-1)}$, $\frac{1}{2}<q<1$;
\item[(ii)]
if $\phi \ne 0$, $\phi <f$, with $\widehat{\eta}_i^j(k)\sim \mathrm{Lap}({c}\,{q}^k)$, for all $i,j$ and $k\geq 0$, then $\varepsilon$-differential privacy is guaranteed for non-faulty agents, with $\varepsilon=\bar{\varepsilon} + {\bar\delta} \frac{f\, d_{\max}^{\mathrm{out}}\, q}{c\,(q-\lambda)}$.
\end{enumerate}
\end{proposition}
\begin{pf*}{Proof.}
The proof follows similar steps and notation to those presented in \citep{nozari2017differentially,huang2012differentially}.
Essentially, equation \eqref{eq:def_diff_priv} is proved by establishing a bijective correspondence between any two executions $(\theta_0^{(1)}, \bm{\eta}^{(1)})$ and $(\theta_0^{(2)}, \bm{\eta}^{(2)})$ that generate the same observations (that is, such that $X_{\theta_0^{(1)}}(\bm{\eta}^{(1)})=X_{\theta_0^{(2)}}(\bm{\eta}^{(2)})$) and then by evaluating the probability by integrating over all the possible executions starting from the same $\delta$-adjacent initial conditions. Without loss of generality, assume that the agents from $1$ to $n-\phi$ are non-faulty, and that the agents from $(n-\phi+1)$ through $n$ are faulty. At a generic time instant $h$ the vector noise generated by all the agents is given as
$
\eta=\big[\eta_1 \ \dots \ \eta_{n-\phi} \ \widehat{\eta}_{n-\phi+1}^T \ \dots \ \widehat{\eta}_{n}^T\big]^T.
$
That is, the first $n-\phi$ components of $\eta$ are the $\eta_i\in\mathbb{R}$ generated by non-faulty agents, while $\widehat{\eta}_i=\big[\widehat{\eta}_i^1 \ \dots \ \widehat{\eta}_i^{d_i^{\mathrm{out}}} \big]^T$, $i\in\sF$, is the noise generated by the $i$-th faulty agent and received by each of its $d_i^{\mathrm{out}}$ out-neighbors (recall that faulty agents can send a different noise signal to each of their out-neighbors). Moreover, note that the vector $\eta\in\R^m$, where $m=(n-\phi)+\sum_{i=1}^\phi d_i^{\mathrm{out}}$. Consider now any pair of $\delta$-adjacent initial conditions $\theta_0^{(1)}$ and $\theta_0^{(2)}$ of the non-faulty agents and an arbitrary set $\mathcal{O}\subset (\mathbb{R}^m)^\mathbb{N}$ of observations. For any $k\geq 0$, let $R_k^{(\ell)}=\{ \bm{\eta}_k \in \Omega_k | X_{k,\theta_0^{(\ell)}} (\bm{\eta}_k) \in \mathcal{O}_k  \}$, $\ell=1,2$, where $\Omega_k=\mathbb{R}^{m(k+1)}$ being the sample space up to time $k$, and $\mathcal{O}_k\subseteq\mathbb{R}^{m(k+1)}$ being the set of observations obtained by truncating the elements of $\mathcal{O}$ to finite subsequences of length $k+1$. Hence, by the continuity of probability \citep[Theorem 1.1.1.iv]{durrett2010probability}
\begin{equation*}
\sP\{ \bm{\eta} \in \Omega | X_{\theta_0^{(\ell)}} (\bm{\eta}_k) \in \mathcal{O} \} = \lim_{k\to\infty} \int_{R_k^{(\ell)}} \mathsf{f}_{m(k+1)}(\bm{\eta}_k^{(\ell)} ) d\bm{\eta}_k^{(\ell)},
\end{equation*}
for $\ell=1,2$, where $\mathsf{f}_{m(k+1)}$ is the $m(k+1)$-dimensional joint probability distribution given as
\begin{equation}
\label{eq:joint_pdf}
\begin{split}
& \mathsf{f}_{m(k+1)}  (\bm{\eta}_k) \\
& = \prod\limits_{h=0}^k \bigg[ \prod\limits_{i=1}^{n-\phi} \mathcal{L}(\eta_i(h);b(h))  \cdot  \!\!\!\!\!\! \prod\limits_{i=n-\phi+1}^n \prod\limits_{j=1}^{d_i^{\mathrm{out}}} \mathcal{L}(\widehat{\eta}_i^j(h); \widehat{b}(h))  \bigg]\\
& =  \prod\limits_{h=0}^k \prod\limits_{i=1}^{n-\phi} \mathcal{L}(\eta_i(h);b(h))\cdot \prod\limits_{h=0}^k \prod\limits_{i=n-\phi+1}^n \prod\limits_{j=1}^{d_i^{\mathrm{out}}} \mathcal{L}(\widehat{\eta}_i^j(h); \widehat{b}(h)) \\ 
& = \mathsf{f}_{(n-\phi)(k+1)}(\bm{\eta}_k) \cdot \widehat{\mathsf{f}}_{(m-n+\phi)(k+1)}(\bm{\eta}_k).
\end{split}
\end{equation}
Now, without loss of generality, assume that $\theta_{0,i_0}^{(2)}=\theta_{0,i_0}^{(1)}+\delta_1$ for some $i_0\in\{1,\dots,n-\phi\}$, where $0\leq\delta_1 \leq \delta$, and $\theta_{0,i}^{(2)}=\theta_{0,i}^{(1)}$ for all $i\neq i_0$. Then, for any $\bm{\eta}_k^{(1)}\in R_k^{(1)}$, define $\bm{\eta}_k^{(2)}$ as
\begin{equation}
\label{eq:privacy_noise}
\eta_i^{(2)}(h)=
\begin{cases}
\eta_{i}^{(1)}(h)-a_{i}^h\, \delta_1, & \mbox{if } i=i_0,\\
\eta_i^{(1)}(h), & \mbox{if } i\neq i_0,
\end{cases}
\end{equation}
for $h\in\{0,\dots , k\}$.
Then, from \eqref{eq:adding_noise}, for $i=i_0$ we have $x_{i_0}^{(2)}(0)=\theta_{0,i_0}^{(2)}+\eta_{i_0}^{(2)}(0)=\theta_{0,i_0}^{(1)}+\delta_1+\eta_{i_0}^{(1)}(0)-\delta_1=x_{i_0}^{(1)}(0)$, while for $i\neq i_0$ we have $x_{i}^{(2)}(0)=\theta_{0,i}^{(2)}+\eta_{i}^{(2)}(0)=\theta_{0,i}^{(1)}+\eta_{i}^{(1)}(0)=x_i^{(1)}(0)$. That is, \eqref{eq:privacy_noise} implies $x^{(2)}(0)=x^{(1)}(0)$. Since Algorithm \ref{alg:DP-MSR} performs only deterministic operations on the data that receives as an input (in fact, it first sorts the values and then cuts the smallest $f$ and largest $f$ values), then $x^{(2)}(0)=x^{(1)}(0)$ implies $\mathcal{N}_i^{\ast,(2)}(0)=\mathcal{N}_i^{\ast,(1)}(0)$, for $i\in\{1,\dots, n-\phi\}$ (i.e. the sets $\sN_i^{\ast}$ generated from the two $\delta$-adjacent conditions are the same, because in both cases the same values are removed by the algorithm). Therefore, by induction, it can be easily shown that from \eqref{eq:privacy_noise} we get $\theta_i^{(2)}(h)=\theta_{i}^{(1)}(h)+a_{i}^h\, \delta_1$, if $i=i_0$, and $\theta_i^{(2)}(h)=\theta_i^{(1)}(h)$, if $i\neq i_0$,
and also $x^{(2)}(h)=x^{(1)}(h)$, which implies $\mathcal{N}_i^{\ast,(2)}(h)=\mathcal{N}_i^{\ast,(1)}(h)$, for $i\in\{1,\dots, n-\phi\}$ and $h\in\{0,\dots , k\}$. That is, $X_{k,\theta_0^{(1)}}(\bm{\eta}_k^{(1)})=X_{k,\theta_0^{(2)}}(\bm{\eta}_k^{(2)})$, so $\bm{\eta}^{(2)}\in R_k^{(2)}$, and this defines a bijective correspondence between the two executions.
Hence, for any $\bm{\eta}_k^{(2)}\in R_k^{(2)}$ there exists a unique $(\bm{\eta}_k^{(1)}, \Delta\bm{\eta}_k)\in R_k^{(1)} \times \mathbb{R}^{(n-\phi)(k+1)}$ such that 
$
\bm{\eta}_k^{(2)}=\bm{\eta}_k^{(1)}+\Delta \bm{\eta}_k,
$
where $\Delta \bm{\eta}_k$ is fixed. Therefore, using a change of variable as in \citep{nozari2017differentially}, we get $\sP\{ \bm{\eta}\in\Omega | X_{\theta_0^{(2)}}(\bm{\eta}) \in\mathcal{O} \} = \lim_{k\to\infty} \! \int_{R_k^{(1)}} \!\! \mathsf{f}_{m(k+1)}(\bm{\eta}_k^{(1)} + \Delta \bm{\eta}_k ) d\bm{\eta}_k^{(1)}$.\\
Now, we first prove statement (i) of the proposition (i.e. $\phi = 0$) and then we prove statement (ii).\\
\noindent {\em Proof of statement (i):} In this case $\phi=0$. This implies that $\mathsf{f}_{m(k+1)}  (\bm{\eta}_k)$ defined in \eqref{eq:joint_pdf} is simply equal to  $\mathsf{f}_{(n-\phi)(k+1)}(\bm{\eta}_k)$. Hence, the relation between the probability of any subsequence of an individual execution $\bm{\eta}_k^{(1)}$ and its corresponding execution $\bm{\eta}_k^{(2)}$ for a particular observation is given by
\begin{equation}
\label{eq:ratio_epsilon_faultfree}
\begin{split}
&\frac{\mathsf{f}_{(n-\phi)(k+1)}(\bm{\eta}_k^{(1)})}{\mathsf{f}_{(n-\phi)(k+1)}(\bm{\eta}_k^{(1)} \! + \! \Delta\bm{\eta}_k)}  = \frac{ \prod\limits_{h=0}^k \prod\limits_{i=1}^{n-\phi} \mathcal{L}(\eta_i^{(1)}(h); b(h))}{\prod\limits_{h=0}^k \prod\limits_{i=1}^{n-\phi} \mathcal{L}(\eta_i^{(1)}(h) \! + \! \Delta \eta_i(h); b(h))}\\
&\qquad = \frac{\prod\limits_{h=0}^k \mathcal{L}(\eta_{i_0}^{(1)}(h); b(h))}{\prod\limits_{h=0}^k \mathcal{L}(\eta_{i_0}^{(1)}(h) \! + \! \Delta \eta_{i_0}(h); b(h))}
\leq  \prod\limits_{h=0}^k e^{\frac{\lvert \Delta \eta_{i_0}(h) \rvert}{b(h)}} \\
&\qquad \leq e^{ \sum\limits_{h=0}^k \frac{a_{i_0}^h \delta}{c q^h}  }
\leq  e^{ \frac{\delta}{c} \sum\limits_{h=0}^k \left(\frac{1}{2 q}\right)^h }
\end{split}
\end{equation}
where the last step follows from $a_i<1/2$. Thus, we have 
\begin{equation}\label{eqn:proof}
\mathsf{f}_{(n-\phi)(k+1)}(\bm{\eta}_k^{(1)}) \leq e^{ \frac{\delta}{c} \sum\limits_{h=0}^k \left(\frac{1}{2 q}\right)^h } \mathsf{f}_{(n-\phi)(k+1)}(\bm{\eta}_k^{(1)} \! + \! \Delta \bm{\eta}_k).
\end{equation}
Therefore, integrating both sides of \eqref{eqn:proof} over all the executions  $R_k^{(1)}$, taking the limit for $k\to\infty$ and considering that the geometric series in the exponent is convergent since from the hypotheses $\frac{1}{2}<q<1$, we have 
$$
\sP\{\bm{\eta}\in\Omega | X_{\theta_0^{(1)}}(\bm{\eta}) \in\mathcal{O} \} \leq e^{\delta \frac{2q}{c\,(2q-1)}} \sP\{\bm{\eta}\in\Omega | X_{\theta_0^{(2)}}(\bm{\eta}) \in\mathcal{O} \}.
$$
The result is then proved by definition of differential privacy.\\
\noindent {\em Proof of statement (ii):} Now, in this case $\phi>0$. For any $\bm{\eta}_k^{(1)}$ define
\begin{equation}
\label{eq:privacy_noise_attack}
\widehat{\eta}_i^{j,(2)}(h)=\widehat{\eta}_i^{j,(1)}(h)-\widehat{\delta}_i^j(h),
\end{equation}
for $h=\{0, \dots, k\}$. Similarly to what has been previously done, it can be shown that $x_i^{j,(2)}(0)=\widehat{\theta}_i^{j,(2)}(0)+\widehat{\eta}_i^{j,(2)}(0)=\widehat{\theta}_i^{j,(1)}(0)+\widehat{\delta}_i^j(0)+\widehat{\eta}_i^{j,(1)}(0)-\widehat{\delta}_i^j(0)=x_i^{j,(1)}(0)$. Therefore, also in this case, due to the deterministic operations carried out by Algorithm \ref{alg:DP-MSR} on its input data, we have that $\mathcal{N}_i^{\ast,(2)}(0)=\mathcal{N}_i^{\ast,(1)}(0)$, for $i=\{1,\dots,n-\phi\}$, and hence, from \eqref{eq:privacy_noise} and \eqref{eq:privacy_noise_attack}, by induction, we have again $X_{k,\theta_0^{(1)}}(\bm{\eta}_k^{(1)})=X_{k,\theta_0^{(2)}}(\bm{\eta}_k^{(2)})$.
Thus, from \eqref{eq:joint_pdf}, the relation between probabilities of any of these two corresponding executions is given by
\begin{equation}\label{eqn:relations}
\frac{\mathsf{f}_{m(k+1)}(\bm{\eta}_k^{(1)})}{\mathsf{f}_{m(k+1)}(\bm{\eta}_k^{(2)})} = \frac{\mathsf{f}_{(n-\phi)(k+1)}(\bm{\eta}_k^{(1)})}{\mathsf{f}_{(n-\phi)(k+1)}(\bm{\eta}_k^{(2)})} \,\cdot \, \frac{\widehat{\mathsf{f}}_{(m-n+\phi)(k+1)}(\bm{\eta}_k^{(1)})}{\widehat{\mathsf{f}}_{(m-n+\phi)(k+1)}(\bm{\eta}_k^{(2)})},
\end{equation}
where the first term is given in \eqref{eq:ratio_epsilon_faultfree} and the second one, omitting $(m-n+\phi)(k+1)$ in the subscripts, is
\begin{equation*}
\label{eq:ratio_epsilon_faulty}
\begin{split}
&\frac{\widehat{\mathsf{f}}(\bm{\eta}_k^{(1)})}{\widehat{\mathsf{f}}(\bm{\eta}_k^{(1)} \! + \! \Delta\bm{\eta}_k)}  = \frac{ \prod\limits_{h=0}^k \prod\limits_{i=n-\phi+1}^{n} \prod\limits_{j=1}^{d_i^{\mathrm{out}}} \mathcal{L}(\widehat{\eta}_i^{j,(1)}(h); \widehat{b}(h))}{\prod\limits_{h=0}^k \prod\limits_{i=n-\phi+1}^{n} \prod\limits_{j=1}^{d_i^{\mathrm{out}}} \mathcal{L}(\widehat{\eta}_i^{j,(1)}(h)  \! + \! \Delta \widehat{\eta}_i^j(h); \widehat{b}(h))}\\
& \leq \!  \prod\limits_{h=0}^k \prod\limits_{i=n-\phi+1}^{n} \prod\limits_{j=1}^{d_i^{\mathrm{out}}} e^{\frac{\lvert \Delta \hat{\eta}_i^j(h) \rvert}{\hat{b}(h)}}
\! \leq \! \prod\limits_{h=0}^k e^{ \frac{\phi d_{\max}^{\mathrm{out}} {\bar\delta} \lambda^h}{c q^h}} 
\! \leq \! e^{ \frac{\bar\delta f d_{\max}^{\mathrm{out} }}{c}  \sum\limits_{h=0}^k \left(\frac{\lambda }{q}\right)^h }
\end{split}
\end{equation*}
where we used the bound in \eqref{eq:attack_bound}.
Combining the above two relations and \eqref{eq:ratio_epsilon_faultfree} we finally get 
$$
\mathsf{f}_{m(k+1)}(\bm{\eta}_k^{(1)}) \leq e^{ \frac{\delta}{c} \! \sum\limits_{h=0}^k \left(\frac{1}{2 q}\right)^h + \frac{\bar\delta f d_{\max}^{\mathrm{out} }}{c} \! \sum\limits_{h=0}^k \left(\frac{\lambda }{q}\right)^h} \!\!\! \mathsf{f}_{m(k+1)}(\bm{\eta}_k^{(1)} +  \Delta \bm{\eta}_k).
$$
In the same way as before, integrating over all the possible executions, using a change of variables and taking the limit, we finally get 
$$
\sP\{\bm{\eta}\in\Omega  | X_{\theta_0^{(1)}}(\bm{\eta}) \in\mathcal{O} \} \leq e^{\bar{\varepsilon} + \bar\delta \frac{f\, d_{\max}^{\mathrm{out}} q}{c\,(q-\lambda)}} \ \sP\{\bm{\eta}\in\Omega | X_{\theta_0^{(2)}}(\bm{\eta}) \in\mathcal{O} \},
$$
that establishes the desired result by definition of differential privacy. \hfill$\square$
\end{pf*}
Note that, if the convergence to zero of $\widehat{\delta}_i^j(k)$ in \eqref{eq:attack_bound} is in one-step, i.e. $\lambda=0$, then $\varepsilon=\bar{\varepsilon} + {\bar\delta} \frac{f\, d_{\max}^{\mathrm{out}}}{c}$.
Moreover, if $\bar{\delta}= 0$, i.e. the faulty agents are not completely omniscient, then Proposition \ref{prop:privacy} implies that  faulty agents do not perturb the differential privacy of the multi-agent system. 
\begin{remark}\label{rem:4}
The major difference of Proposition \ref{prop:privacy} in the case $\phi=0$ with respect to the results in \citep{nozari2017differentially} is that we consider directed graphs whose topology changes over time, due to the operations performed by Algorithm \ref{alg:DP-MSR}. With respect to this, we note that the only feature of Algorithm \ref{alg:DP-MSR} used in the previous proof is that the algorithm performs deterministic operations on its input data. Hence, \emph{the proof is valid for any deterministic algorithm}. 
\end{remark}
\section{Simulation example}
\label{sec:simulations}
\begin{figure}[thbp]
\centering
\includegraphics[width=0.4\textwidth]{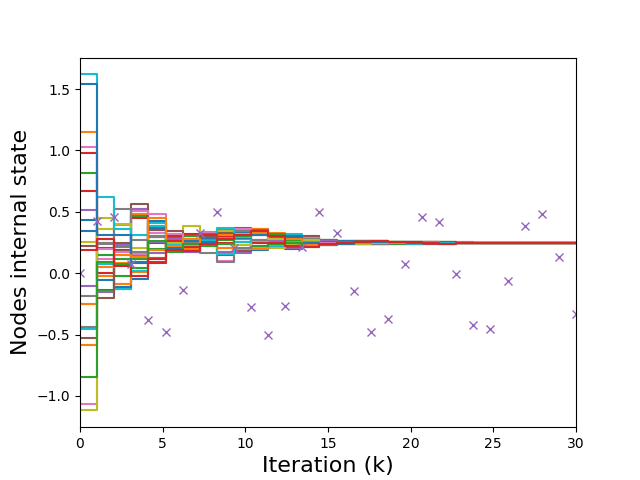}
\caption{Time evolution of the internal states of non-faulty agents for the multi-agent system in Section \ref{sec:simulations}. The time evolution of the attack signal sent by the faulty agent is denoted by a cross.}\label{fig:evolution}
\end{figure}

\begin{figure}[thbp]
\centering
\includegraphics[width=0.4\textwidth]{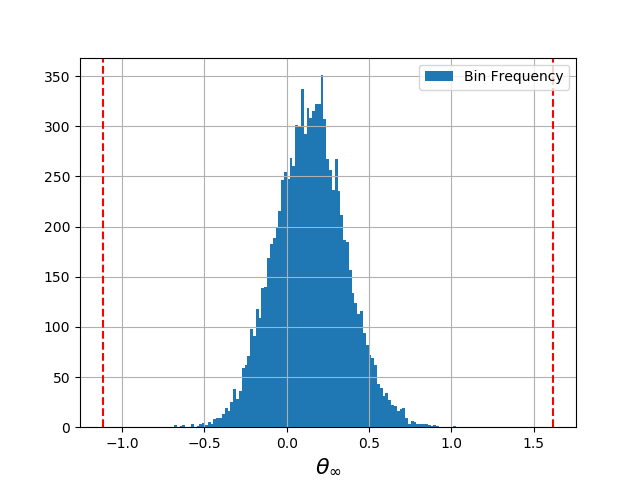}
\caption{Statistical distribution of $\theta_{\infty}$. Dashed lines indicate the smallest and the largest initial conditions of non-faulty agents.}\label{fig:distribution}
\end{figure}

We now report simulation results for a multi-agent system of $n = 25$ agents having $f = 1$ faulty agent (say, agent $1$). The agents interact over a circulant digraph where each agent communicates with $8$ agents ahead. It can be shown that such a graph is $4$-robust, see \citep[Theorem 1]{Use_Pan_18}. Initial conditions of the non-faulty agents are taken from the standard normal distribution and are kept the same across the simulations. At each time-step $k$, non-faulty agents update their internal state and send messages to their out-neighbors in accordance to the DP-MSR, while the faulty agent sends the message $\hat x_1^j(k) = 0.5\sin(k) + \eta_1^j(k)$, where $j\in \sN_1^{out}$ and $\eta_1^j(k) \sim \mathrm{Lap}(\widehat{c}\,\widehat{q}^k)$, with $\widehat q = 0.9$ and $\widehat c = 0.8$. 
We set the parameters in Algorithm \ref{alg:DP-MSR} to $c = 1$ and $q = 0.75$. 
Since the graph is $4$-robust, then the DP-MSR algorithm is correct (Proposition \ref{prop:correctness}) and the system reaches mean square consensus to some value of the random variable $\theta_\infty$, whose expected value $\mathbb{E}[\theta_\infty]$ is contained between the smallest and the largest values of the initial conditions of the non-faulty agents. This prediction is confirmed in Figure \ref{fig:evolution}. In Figure \ref{fig:distribution}, the statistical distribution of $\theta_{\infty}$ is shown, which has been obtained by running $10^4$ simulations. Further, we numerically found that $\mathrm{var}\left( \theta_\infty \right) \approx 0.05$, in accordance with the theoretical bound defined in \eqref{eq:variance_bounds}, that is $0.0037\leq \mathrm{var}(\theta_\infty)\leq 27.4286$.

\section{Conclusions}
We considered  multi-agent systems interacting over a directed network topology, where a subset of agents is faulty and where the non-faulty agents have the goal of achieving consensus and are subject to an additional differential privacy requirement. Our main contribution is the introduction an algorithm, the DP-MSR algorithm, and the characterization of its correctness, accuracy and differential privacy properties. 
Simulations were used to illustrate the results.
Future work will be aimed at extending the results presented in this paper to consider: (i) the analysis of differential privacy in more general scenarios, where the faulty agents do not respect the network communication policy; (ii) asynchronous systems; (iii) the design of algorithms to detect and isolate faulty agents.
%
%
\begin{ack}                               
The authors wish to thank Prof. Mario di Bernardo for reading this manuscript and for the comments he provided, Dr. Pietro De Lellis, Dr. Noise Holohan and Dr. Andrea Simonetto for the useful discussions on differential privacy and stochastic processes. The authors are also grateful to the anonymous reviewers and the AE for their constructive feedback.
\end{ack}
%
\bibliographystyle{elsarticle-harv}        
\bibliography{refs}           
%
\end{document}